\documentclass[12pt,preprint]{aastex}

\newcommand{\eb}{\begin{equation}}
\newcommand{\ee}{\end{equation}}

\newcommand{\ms}{m~s$^{-1}$}
\newcommand{\uas}{$\mu$as}

\shorttitle{Solar jitter}
\shortauthors{Makarov et al.}
\begin{document}

\title{ASTROMETRIC JITTER OF THE SUN AS A STAR} 
\author{V.V. Makarov\altaffilmark{1}, D. Parker\altaffilmark{2}, R.K. Ulrich \altaffilmark{2}}
\affil{$^1$NASA Exoplanet Science Institute, Caltech, Pasadena, CA 91125}
\affil{$^2$Department of Physics and Astronomy, University of California, Los Angeles, CA 90095}
\email{vvm@caltech.edu}

\begin{abstract}
The daily variation of the solar photocenter over some 11 years is derived
from the Mount Wilson data reprocessed by Ulrich et al. 2010 to closely
match the surface distribution of solar irradiance. The standard deviations
of astrometric jitter are 0.52 $\mu$AU and 0.39 $\mu$AU in the equatorial
and the axial dimensions, respectively. The overall dispersion is strongly correlated
with the solar cycle, reaching $0.91~\mu$AU at the maximum activity in 2000.
The largest short-term deviations from the running average (up to 2.6 $\mu$AU)
occur when a group of large spots happen to lie on one side with respect to
the center of the disk. The amplitude spectrum of the photocenter variations
never exceeds $0.033$ $\mu$AU for the range of periods $0.6$--$1.4$ yr,
corresponding to the orbital periods of planets in the habitable zone. Astrometric detection of Earth-like planets around
stars as quiet as the Sun is not affected by star spot noise, but the prospects
for more active stars may be limited to giant planets.
\end{abstract}

\section{Introduction}
\label{firstpage}

The prospects of finding habitable planets orbiting nearby solar-type
stars are to a large degree associated with the ultra-precise astrometric instruments
under development or construction, such as the SIM Observatory \citep{sha,unw,catpa}
and Gaia \citep{cas}. The Earth orbiting
the Sun produces an observable astrometric wobble of 3 $\mu$AU (micro-AU)
and a radial velocity variation of 0.089 \ms, if seen equator-on (inclination
$i=90\degr$). A pole-on
configuration ($i=0\degr$ or $180\degr$) is optimal for astrometry, because 
the reflex motion signal is
present in both dimensions of the sky projection, whereas the radial velocity
amplitude, which is proportional to $\sin i$, drops to zero.
For a high signal-to-noise (S/N) ratio detection, the total error budget
of the prospective exoplanet detection techniques should be well below
these values. The astrophysical jitter caused by the rotation of magnetic
features on the surface (spots, faculae) potentially becomes a significant part of
the total observational error in this domain of accuracy, even for fairly
common inactive dwarfs. The recently completed double blind test of
planet-detection capabilities with the SIM Observatory produced
encouraging results even for complex multiple systems \citep{tra},
but magnetic jitter was not taken into account in those simulations.
Semi-analytical considerations in \citep{mak09}
supported by Monte-Carlo simulations and indirect observational evidence
determined that the astrometric method is much less sensitive to the effects
of magnetic features, at least by an order of magnitude for solar-type stars,
than the Doppler shift technique. Similar conclusions were drawn by \citet{cat}
for the Sun, based on a sophisticated model of sunspot activity and extensive
numerical simulation. 

The variable distribution of surface brightness is the cause 
of both the astrometric jitter and the total flux variation. Therefore, the former
can be estimated from the latter, given a model describing the
number of spots (or bright features), their size and lifetimes. The dispersion
of the total solar irradiance (TSI) observed for a few decades with a number of
satellites \citep{fro} sets the standard of quiet stars ($4\cdot 10^{-4}$,
or 400 ppm in normalized flux). The light curves from Kepler for some
120000 stars revealed that at least half of all solar-type dwarfs in the
field are at least as quiet as the Sun \citep{bas}, displaying lower levels
of photometric dispersion. By extrapolation, about half of nearby dwarfs
should be as amenable to exoplanet detection as the Sun. The aim of this
paper is to determine the solar astrometric jitter directly from observations.
The results can be used to verify and correct the existing models of
spot activity for other stars.

\section{The data}
\label{data.sec}
We made use of the set of 2881 images of the solar disk reconstructed from Mount Wilson
Observatory magnetogram and intensity ratio images for the period 1996 to 2007
\citep{ulr}. As described in that paper, two options of the AutoClass analysis
yielded an 18 and a 37 class set of solar surface features. The original Mount Wilson images
in the intensity ratio $I_{\lambda5250}/I_{\lambda5237}$ and magnetic field strength
$|B_{\lambda5250}|$ were expanded in the corresponding sets of classes and calibrated
against the TSI data \citep{fro}. The resulting solar
irradiance images have the full resolution of the original data, with a value of normalized TSI assigned
to each pixel. Essentially, these images are two-dimensional
maps of bolometric surface intensity. The reconstructed images are only 95\% of the
full solar disk in radius. The units of pixel values are W$\cdot$m$^{-2}$, and they are
normalized to unit total area. For this study, we used the 37-class reconstruction,
which provides a slightly better match to the TSI, although this improvement is
not important for our purpose. \citet{ulr} discuss the "ring effect" in the reconstructed TSI
maps, which manifests itself as a kind of diffraction ring surrounding high-contrast,
compact features, especially dark spots. The spatial scale of these artefacts is small,
and the impact on the estimated magnetic jitter is negligible.

The processing of the TSI maps was straightforward. Each of the 2881 images was integrated
in first moment with respect to a fixed pixel row, e.g.,
\eb
\Delta x= \sum_{i,j} I(x_{ij},y_{ij})\, x_{ij}/\sum_{i,j} I(x_{ij},y_{ij}),
\ee
and similarly for $\Delta y$. The $x$ axis is aligned with the solar equator,
and the $y$ axis with the rotation axis. The derived offsets were brought to zero mean and
converted to $\mu$AU. Since the TSI images are based on the observed distributions of
magnetic strength and the ratio of two monochromatic intensities, they are almost
devoid of the limb darkening effects. Therefore, we do not apply any limb darkening corrections
as the counter-acting effects of the broad-band intensity reduction toward the
limb and the increasing  contrast of faculae with respect to the local photosphere
\citep{fouk} have been balanced in the images by construction.

\section{Astrometric jitter from the magnetic activity}
\begin{figure}[htbp]
\plotone{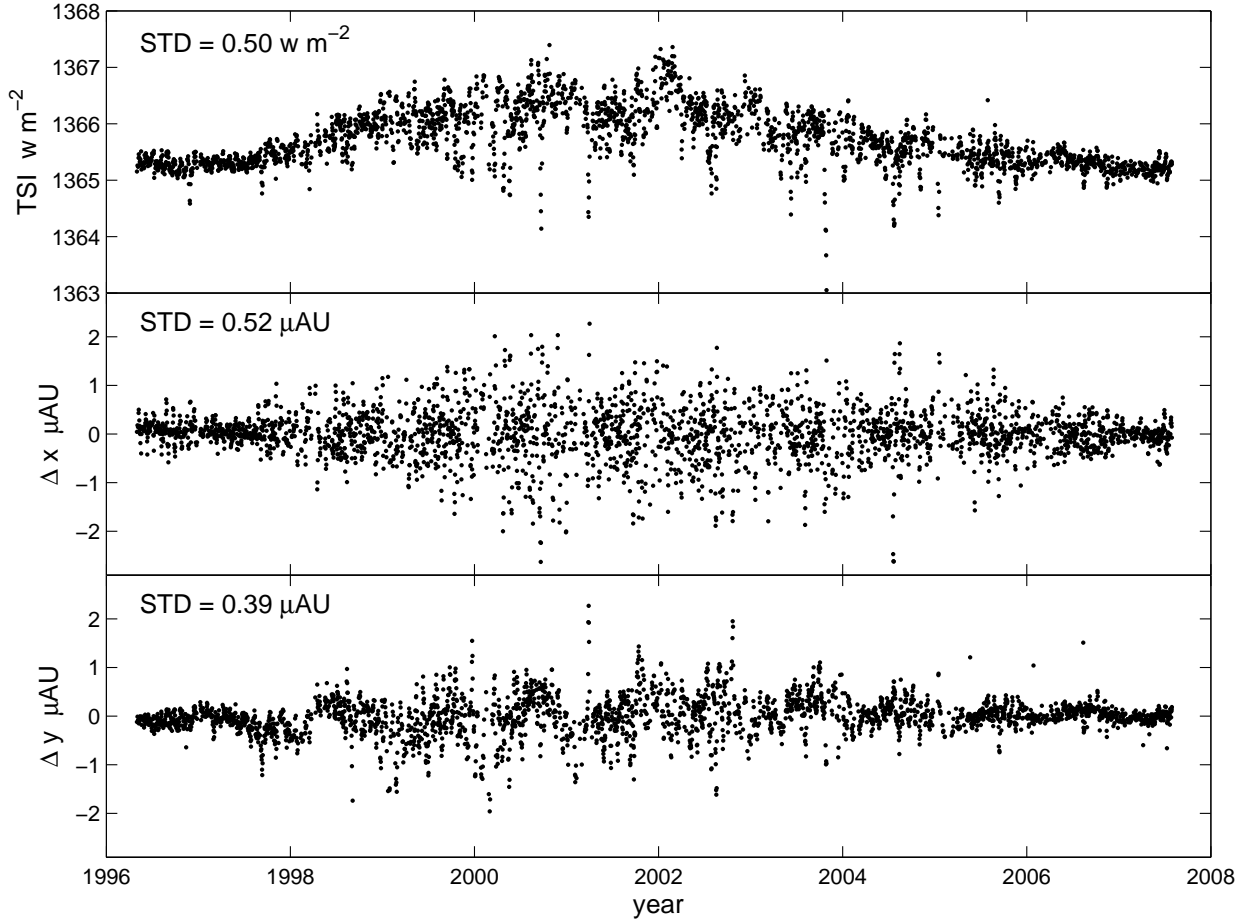}
\caption{Daily variations of the total solar irradiance, and the photocenter position
in the equatorial ($\Delta x$) and axial ($\Delta y$) dimensions, derived from the
reconstructed TSI maps.} 
  \label{xy.fig}
\end{figure}

The main results of this study are shown in Fig.~\ref{xy.fig}. The top panel displaying
the integrated TSI is shown only for reference, because the TSI images were constructed
to match the space-based TSI measurements as close as possible, thus, the
plot does not contain any new information. The middle and bottom panels show the
daily photocenter offsets of the solar disk over some 11 years in the equatorial and
axial dimensions, respectively. The TSI has been  for some time known to vary systematically with
the solar cycle \citep{wil}, but the exact contribution of the opposing
effects (faculae and spots) to this process is still somewhat controversial. The Sun
is brighter when its level of magnetic activity is elevated, while younger and more active
stars tend to display the opposite correlation \citep{rad}. The peak-to-peak variation of TSI is approximately 2 W$\cdot$m$^{-2}$, or 0.15\%. The maximum of activity in 2000--2002 is marked with
a brighter average and a much larger dispersion of TSI. A running average standard deviation
over 3 months varies between $0.14$ $\mu$AU and $0.91$ $\mu$AU in $\Delta x$. The
overall standard deviations are $0.52$ $\mu$AU for $\Delta x$ and $0.39$ $\mu$AU for 
$\Delta y$. Predictably, the dispersion of $\Delta y$ is smaller, because major spots
occur in the equatorial zone of the disk.

The distinct variation in $\Delta y$ with a period of 1 year is probably an artefact related
to the seasonal change of the solar obliquity. In September, the solar axis is tilted by
$7.25\degr$, with its north pole toward the Earth. Since the distribution of magnetic
features is symmetric about the solar equator, more spots and faculae appear in the southern
part of the disk than in the northern part, as seen by a ground-based observer. This 1-year
modulation of the $\Delta y$ offset is germane to the Sun and will not be observed
on other stars. A periodogram analysis determined that the best fitting sinusoid
has a period of 1 yr and an amplitude of $0.11$ $\mu$AU.

\begin{figure}[htbp]
\plotone{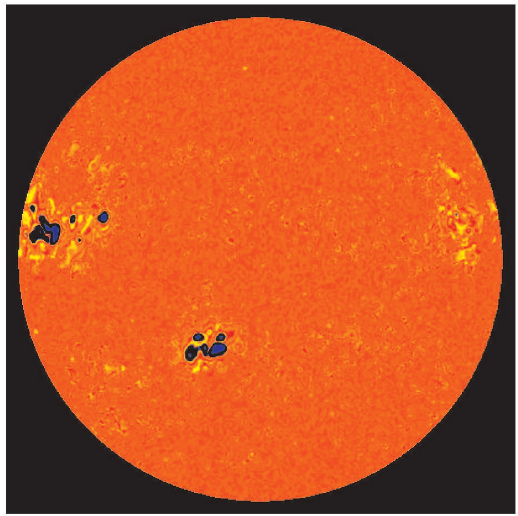}
\caption{Solar surface irradiance distribution on 2000.09.19, color-coded to
emphasize the dark spots and the bright plage areas. North is up, east is
left. An exceptionally large deviation of $-2.6$ $\mu$AU along
the equator would be observed astrometrically.} 
  \label{sun.fig}
\end{figure}

The running average of the $\Delta x$ perturbations is not correlated with
the solar cycle, because the probability of magnetic features to be located in 
the eastern or western hemispheres is the same. The scatter of daily variation
is, however, strongly correlated with the level of magnetic activity. The distribution
of TSI perturbations around the running mean is asymmetric, with a longer tail extending
toward fainter intensities. These short-lived dips in TSI are caused by particularly
large sunspot groups, and normally happen at times of elevated activity. The distribution of
astrometric perturbations around the running mean are fairly symmetric, as expected.
The largest astrometric perturbation $\Delta x$ took place on 2000.09.19, when a distant
observer would have detected a westward photocenter shift of $2.6$ $\mu$AU. Fig.~\ref{sun.fig}
depicts the distribution of surface brightness of the Sun on that date. The color scale
is arbitrary, adjusted in such a way that the dark spots (black and blue) and the
faculae (yellow) can be clearly seen. This unusually large perturbation was caused
by two sunspot groups, which happened to be in the eastern hemisphere, and a bright
active region in the opposite western hemisphere.

The temporal variations of solar surface brightness are commonly considered
to be caused by small- and medium-scale magnetic features, such as photospheric
spots and plages. Is it possible that some other, non-magnetic structures are
present in the surface brightness distribution, which would not be captured
by the Mount Wilson magnetograms that served as the original data for the TSI
reconstruction? \citet{ulr} found that the reconstructed irradiance followed
the actual TSI variations, observed with Virgo, to a correlation coefficient of
0.9625. This corresponds to a standard deviation of 28\% relative to the
variation of TSI, if the error introduced by the proxy reconstruction is
uncorrelated with the true light curve. Although it is not possible to tell
how this error is distributed across the solar disk, we condend that the true
astrometric jitter can not be {\it larger} than our estimates by more than this amount,
which is 4\% when added in quadrature.
The error is carried by many classes of different spatial scale, but
only large scale perturbations count in the integration in first moment.
Most of this error is probably confined to small-scale structures,
which should cancel in integration. Known imperfections, such as the ringing features
centered on the centers of activity, caused by too many plage classes, do not
result in an appreciable image shift. Besides, some of the TSI variation not accounted 
for by the reconstruction is due to magnetic features appearing in the outer rim of the solar
disk (5\% of the radius), which is missing in our data.

\section{Implications for Earth-like planet detection}
Planets revolving around their host stars produce sinusoidal variations in the
astrometric position. The main harmonic of these variations has the orbital
period of the planet, and an amplitude defined by the mass of the star, the
mass of the planet, and the orbital semimajor axis. Planets in eccentric orbits
also produce higher-order harmonics (overtones) of smaller amplitudes.
The spectroscopic method of exoplanet detection, based on precision measurements
of stellar radial velocity, utilizes the Lomb-Scargle periodogram analysis
\citep[e.g.,][]{fis},
which is aimed at detecting statistically significant sinusoidal variations
in irregularly sampled data with a zero mean \citep{sca}. A similar ``joint"
power spectrum periodogram was suggested for 2D astrometric detection
by \citet{cat}. In order to evaluate the
impact of magnetic jitter on detection of low-mass planets, we employ in
this paper a generalized amplitude spectrum analysis, which, unlike the Lomb-Scargle
periodogram, includes the constant term in the set of fitting functions.
A constant offset in position is a physical astrometric parameter, which
can not be simply subtracted from the raw data.

The linear problem
\eb
\left[{\bf 1},\,\cos(\omega{\bf t}),\,\sin(\omega{\bf t})\right]\, {\bf s}
=\Delta{\bf x}
\ee
is solved for a grid of frequencies $\omega=2\pi/p$, where
${\bf t}$ is the column vector of observation times, and  $\Delta{\bf x}$ is
the column vector of $x$-coordinate offsets. If $\tilde{\bf s}$ is the
Least-Squares solution (a 3-vector), the generalized amplitude spectrum as a
function of period is
\eb
D(p)=\sqrt{\tilde s_2^2+\tilde s_3^2}.
\ee
The interpretation of this function is straightforward: it is the amplitude
of the best-fitting sinusoid with the given period $p$ and a free phase.
A planet with an orbital period $p$ can be confidently detected if
the amplitude of its main harmonic is much larger than $D(p)$. It corresponds to
the square root of the ``periodogram power" of the Lomb-Scargle periodogram
commonly used in exoplanet detection.

Fig.~\ref{pow.fig} shows the $D(p)$ spectrum of the $\Delta x$ series
derived from the TSI maps. The calculation was restricted to the range of
habitable zone around the Sun, which corresponds to, somewhat generously,
0.6--1.4 years in orbital period. The spectrum
is distributed non-uniformly, so that white noise would not be
a good model for this kind of perturbation. The largest sinusoidal variation
has a period of $\simeq 1.1$ yr and an amplitude of $0.033$ $\mu$AU.
The Earth generates an astrometric wobble of the Sun of 3 $\mu$AU, whose power
is almost entirely in the main harmonic (1 year) due to the very low eccentricity.
Thus, the Earth can be detected by astrometric means at a very comfortable
S/N of greater than 90, as far as magnetic jitter is concerned, if a similar
observational cadence is achieved. For a
solar-like star at 10 pc, the amplitude spectrum of magnetic jitter is not greater
than $0.0033$ \uas, which is vanishingly small compared to the expected
single measurement precision of SIM ($\ge 1$ \uas) or Gaia ($\ge 8$ \uas).

\section{Discussion}
We determined that the solar jitter caused by magnetic features is
small and should not preclude astrometrists from detecting habitable
planets as small as the Earth. A few additional considerations should be made when
extrapolating this result to other stars or to actual astrometric observations. 

\begin{figure}[htbp]
\plotone{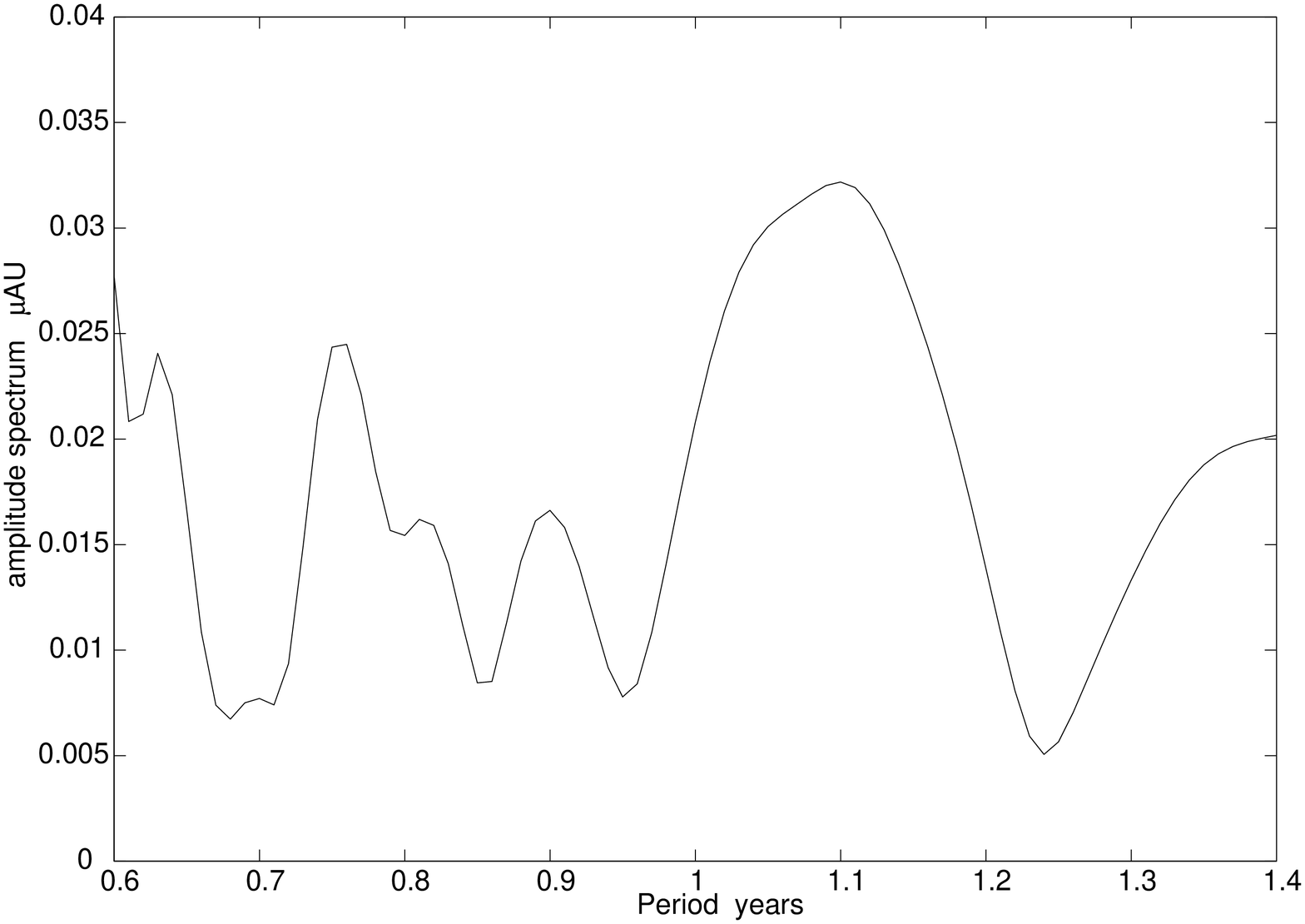}
\caption{Amplitude spectrum of solar astrometric jitter in the
equatorial dimension for the range of periods corresponding to the habitable zone.} 
  \label{pow.fig}
\end{figure}

The ultra-precise photometric measurements with Kepler confirmed that about
half of all solar-type dwarfs are as quiet as the Sun, or more \citep{bas}.
The stars that are more active than the Sun have much larger filling factors,
and their magnetic features may be long-lived (up to a few weeks). Both
photometric and astrometric variability is proportional to the average total area
$A_s$ occupied by spots \citep{mak09}. In the model currently adopted for
planet detection simulations with SIM (to be published elsewhere), 
the total area is related to the index
of chromospheric activity $\log R'_{\rm HK}$ through
\eb
\log A_s=12.468+1.75\log R'_{\rm HK},
\ee
which yields 6137 micro-stellar-hemispheres (MSH) for the Sun. As roughly
half of nearby field stars are more chromospherically active than the Sun
\citep{gra}, they are expected to have larger spot groups and,
therefore, jitter amplitudes. At $\log R'_{\rm HK}=-4.6$, the expected
area and the jitter are 5 times the solar values, and at $\log R'_{\rm HK}=-4.2$,
25 times. This model gives fairly accurate predictions for the active star
$\kappa^1$ Ceti, carefully studied with MOST \citep{wal}. Furthermore,
an additional factor of 2--6 increase may ensue from the limited duration of an astrometric
mission. Both SIM and Gaia will nominally operate for 5 years. This is considerably
shorter than the solar cycle. If a solar analog is observed at the height
of its activity, the level of magnetic jitter may be higher than the
multi-year average. Thus, magnetic jitter can indeed pose the natural limit
to the sensitivity of the astrometric method at stars significantly more
active than the Sun.

The amplitude spectrum periodogram shown in Fig.~\ref{pow.fig} is derived from
a high-frequency observational cadence of 2881 data points over 11 years. We
expect only about 200 2D measurements to be taken for each target star with
SIM Lite during its 5-year nominal mission. If the astrometric jitter is completely
random and uncorrelated, the variance of the amplitude of a random harmonic in the
spectrum is inversely proportional to the number of data points. In that case, the
amplitude spectrum will rise by a factor of $\sqrt{2881/200}=3.8$ for SIM. On the other hand,
if the magnetic activity underlying the astrometric jitter is systematic and deterministic,
the corresponding parts of the spectrum will stay as low is in Fig.~\ref{pow.fig}. Such
systematic variations may occur, for example, if sunspots and plages are not randomly distributed
on the solar surface, but tend to appear in confined areas called ``active longitudes"
in the literature. Currently, there seems to be no strong evidence for the existence
of such long-term structures. A possible north-south asymmetry in the distribution of
magnetic features evolving with the solar cycle is another interesting route of
investigation where the reconstructed TSI images can be used.

The results derived in this paper apply to solar-type stars seen equator-on
(inclination $\simeq 90\degr$). A pole-on configuration is optimal for astrometric
detection, because both dimensions are equally engaged, but it is statistically
less probable. Even at moderate inclinations to the line of sight, the distribution
of magnetic perturbations in $y$-coordinate may be different from that we find for the Sun, if
the magnetic features are confined to the equatorial zone. The features would normally
be seen closer to the side of the limb opposite to the visible pole, producing
a skewed distribution of $\Delta y$ depending on whether the spots or bright areas
are the dominating contributors. However, giant near-polar spots appear to be common
on very active, fast-rotating stars.

\acknowledgments
The research described in this paper was carried out at the Jet Propulsion 
Laboratory, California Institute of Technology, under a contract with the National 
Aeronautics and Space Administration. Observations at the 150-foot tower telescope 
on Mt. Wilson have been
supported over the years by grants from the National Aeronautics and Space
Administration, the National Science Foundation (NSF) and Office of Naval
Research.  Research is currently supported by the NSF through grant
AGS-0958779.

\end{document}